\documentclass{tlp}

\usepackage{subfigure}
\usepackage{amsmath}
\usepackage{graphicx}
\usepackage{color}
\usepackage[latin1]{inputenc}

\usepackage{times}

\usepackage{listings} 
\newenvironment{pcode}
    {\begin{lstlisting}[language=prolog,morekeywords=call,morekeywords=table]}
    {\end{lstlisting}}

\newcounter{mnotei} 
\newcommand{\mnote}[1]{%
 {\scriptsize\textsf{\textcolor{blue}{$^{[\themnotei]}$}}}%
 \marginpar{\scriptsize\textsf{\textcolor{red}{n.\themnotei: #1}}}%
 \stepcounter{mnotei} } 
\renewcommand{\mnote}[1]{}

\newcommand{\compressection}{\vspace{-1em}}
\newcommand{\compressfigure}{\vspace{-1em}}

\begin{document}



\title[]{\vspace{-1em}Parallel Backtracking with Answer Memoing \\
  for Independent And-Parallelism\thanks{
Work partially funded by
EU projects IST-215483 {\em S-Cube} and 
FET IST-231620 {\em HATS}, 
MICINN projects TIN-2008-05624 {\em DOVES},
and CAM project S2009TIC-1465 {\em PROMETIDOS}. Pablo Chico is also
funded by an MICINN FPU 
scholarship.}}


\author[]
       {Pablo Chico de Guzm\'{a}n,$^1$ Amadeo Casas,$^2$ Manuel Carro,$^{1,3}$
        and Manuel V. Hermenegildo$^{1,3}$\\ \ \\
$^1$ School of Computer Science, Univ.\ Polit\'ecnica de Madrid, Spain.\\
     \email{pchico@clip.dia.fi.upm.es, \{mcarro,herme\}@fi.upm.es}\\ 
$^2$ Samsung Research, USA.\\
     \email{amadeo.c@samsung.com}\\ 
$^3$ IMDEA Software Institute, Spain.\\
     \email{\{manuel.carro,manuel.hermenegildo\}@imdea.org}
}






\pagerange{\pageref{firstpage}--\pageref{lastpage}}
\volume{\textbf{10} (3):}
\jdate{July 2011}
\setcounter{page}{1}
\pubyear{2011}

\maketitle

\begin{abstract}
  Goal-level Independent and-parallelism (IAP) is exploited by
  scheduling for simultaneous execution two or more goals which will
  not interfere with each other at run time. This can be done safely
  even if such goals can produce multiple answers.
%
%
  The most successful IAP implementations to date have used
  recomputation of answers and sequentially ordered
  backtracking. While in principle simplifying the implementation,
  recomputation can be very inefficient if the granularity of the
  parallel goals is large enough and they produce several answers,
  while sequentially ordered backtracking limits parallelism.  And,
  despite the expected simplification, the implementation of the
  classic schemes has proved to involve complex engineering, with the
  consequent difficulty for system maintenance and extension, while
  still frequently running into the well-known trapped goal and garbage
  slot problems.
  This work presents an alternative parallel backtracking model for
  IAP and its implementation. The model features parallel out-of-order
  (i.e., non-chronological)
  backtracking and relies on answer memoization to reuse and combine
  answers.
  We show that this approach can bring significant performance
  advantages. Also, it can bring some simplification to the important
  engineering task involved in implementing the backtracking mechanism
  of previous approaches.
\end{abstract}


\begin{keywords}
Parallelism, Logic Programming, Memoization, Backtracking, Performance.
\compressection
\end{keywords}

\compressection
\section{Introduction}
\label{sec:intro}

Widely available multicore processors have brought renewed interest in
languages and tools to efficiently and transparently exploit parallel
execution --- i.e., tools to take care of the difficult~\cite{Karp88}
task of automatically uncovering parallelism in sequential algorithms
and in languages
to succinctly express this parallelism.  These languages
can be used to both write directly parallel applications and as targets
for parallelizing compilers.


Declarative languages (and among them, logic programming languages)
have traditionally been considered attractive for both expressing and
exploiting parallelism due to their clean and simple 
semantics and their expressive power.  A
large amount of work has been done in the area of parallel execution of
logic programs~\cite{partut-toplas}, where two main sources of
parallelism have been exploited: parallelism between goals of a
resolvent (And-Parallelism) and parallelism between the branches 
of the execution (Or-Parallelism).
%
%
Systems efficiently exploiting Or-Parallelism include 
Aurora~\cite{aurora} and MUSE~\cite{muse},
while among those 
exploiting
And-Parallelism,
\&-Prolog~\cite{ngc-and-prolog} and DDAS~\cite{ddas-jlp} are among the
best known ones.  In particular, \&-Prolog exploits \emph{Independent
  And-Parallelism}, where goals to be executed in parallel do not 
compete for bindings to the same variables at run time and are
launched following a nested fork-join structure.
%
Other systems such as ($\&$)ACE~\cite{and-ace-ipps95},
AKL~\cite{sverker-phd}, Andorra-I~\cite{vitor-phd} and the Extended
Andorra Model (EAM)~\cite{eam91,BEAMTPLP2011} have approached a combination of
both or- and and-parallelism. In this paper, we will focus on
independent and-parallelism.

While many IAP implementations
obtained
admirable performance results and achieved efficient memory management, 
implementing synchronization and 
working around problems such as
\emph{trapped goals} (Section~\ref{sec:disordered-backtracking}) and
\emph{garbage slots} in the execution stacks required complex
engineering: extensions to the WAM instruction set, new data
structures,  special stack frames in the stack sets, and
others~\cite{hermenegildo-phd-short}.  Due to this complexity, 
recent approaches have focused instead on simplicity, moving core
components of the implementation to the source level.
In~\cite{hlfullandpar-iclp2008}, a high-level implementation of
goal-level IAP was proposed that showed reasonable speedups despite
the overhead added by the high level of the implementation.
Other recent proposals~\cite{moura08:padl}, with a different focus
than the traditional approaches to parallelism in LP, concentrate
on providing machinery to take advantage of underlying thread-based
OS building blocks.\mnote{We may
  remove this sentence / ref. if we are short in space.}


A critical area in the context of IAP that has also received much
attention is the implementation of backtracking.  Since in IAP by
definition goals do not affect each other, an obvious approach is to
generate all the solutions for these goals in parallel independently,
and then combine them~\cite{Conery-Book-Short}.  However, this
approach has several drawbacks.  First, copying solutions, at least
naively, can imply very significant overhead.  In addition, this
approach can perform an unbounded amount of unnecessary work if, e.g.,
only some of the solutions are actually needed, and it can even be
non-terminating if one of the goals does not fail finitely.
%
For these reasons the operational semantics typically implemented in
IAP systems performs an ordered, right-to-left backtracking. For
example, if execution backtracks into a parallel conjunction such as
\lstinline{a & b & c}, 
the rightmost goal (\lstinline{c}) backtracks first. If it fails, then
\lstinline{b} is backtracked over while \lstinline{c} is recomputed
and so on, until a new solution is found or until the parallel
conjunction fails. 
The advantage of this approach is that it saves memory (since no
solutions need to be copied) and keeps close to the sequential
semantics. However, it also implies that many computations are redone
and a large amount of backtracking work can be essentially sequential.


Herein we propose an improved solution to backtracking in IAP aimed at
reducing recomputation and increasing parallelism while preserving
efficiency.  It combines \emph{memoization of answers to parallel
  goals} (to avoid recomputation), \emph{out-of-order backtracking}
(to exploit parallelism on backtracking), and \emph{incremental
  computation of answers}, to reduce memory consumption and avoid
termination problems. The fact that in this approach the right-to-left
rule may not be followed during parallel backtracking means that
answer generation order can be affected (this of course does not
affect the declarative semantics) but, as explained later, it greatly 
simplifies implementation.
%
%
%
The EAM also supports out-of-order execution of goals. 
However, our approach differs from EAM in 
that the EAM is a more encompassing and complex approach, offering
more parallelism at the cost of more complexity (and overhead) while
our proposal constitutes a simpler and more approachable solution to
implement.

In the following we present our proposal and an IAP implementation of
the approach, and we provide experimental data showing that the amount
of parallelism exploited increases due to the parallelism in backward
execution, while keeping competitive performance for first-answer
queries.  We also observe super-linear speedups, achievable thanks to
memoization of previous answers (which are recomputed in sequential
SLD resolution).\footnote{For brevity we assume some familiarity with
  the WAM~\cite{Warren83,hassan-wamtutorial} and the
  RAP-WAM~\cite{ngc-and-prolog}.}

\compressection
\section{An Overview of IAP with Parallel Backtracking}
\label{sec:high-level-algorithm}

%
In this section we provide a high-level view of the execution
algorithm we propose
to introduce some concepts which we will explain in more detail in
later sections.
%
%

The IAP + parallel backtracking model we propose behaves in many
respects as classical IAP approaches, but it has as its main
difference the use of speculative backward execution (when possible)
to generate additional solutions eagerly.
This brings a number of additional changes which have to be
accommodated. We assume as usual in IAP a number of \emph{agents},
which are normally each attached to their own \emph{stack set},
composed of heap, trail, stack, and goal queue (and often referred in
the following simply as a ``stack''). Active agents are executing code
using their stack set, and they place any new parallel work they find
in their goal queue. Idle agents steal parallel work from the goal
queues of other agents.\footnote{For a more in-depth understanding of
  the memory model and scheduling used in traditional IAP approaches,
  please refer
  to~\cite{ngc-and-prolog,flexmem-europar96,partut-toplas}.}  We will
also assume that stack sets have a new memo area for storing solutions
(explained further later, see Figure~\ref{fig:answer_memo}).

\vspace{-0.5em}
\paragraph{Forward execution:}
as in classical IAP, when a parallel conjunction is first reached, its
goals are started in parallel.  When a goal in the conjunction fails
without returning any solution, the whole conjunction fails. And when
all goals have found a solution, execution proceeds.  However, 
and differently to classical IAP, if a
solution has been found for some goals, but not for all, the agents
which did finish may speculatively perform backward execution for the
goals they executed (unless there is a need for agents to execute work
which is not speculative, e.g., to generate the first answer to a goal).  This in turn brings the need to stash away
the generated solutions in order to continue searching for more
answers (which are also saved).  When all goals find a solution, those
which were speculatively executing are suspended (to preserve the
property of no-slowdown w.r.t.\ sequential
execution~\cite{sinsi-jlp}), their state is saved to be 
resumed later, and their first answer is reinstalled.

\vspace{-0.5em}
\paragraph{Backward execution:}
we only perform backtracking on the goals of a parallel conjunction
which are on top of the stacks.  If necessary, stack sections are
reordered to move trapped goals to the top of the stack.  In order not
to impose a rigid ordering,
we allow backtracking on these goals to
proceed in an arbitrary order (i.e., not necessarily corresponding to
the lexical right-to-left order).  This opens the possibility of
performing backtracking in parallel, which brings some additional
issues to take care of:

\begin{itemize}
\item When some of the goals executing backtracking in parallel
  find a new answer, backtracking stops by suspending the
  rest of the goals and saving their state.
\item The solution found is saved in the memoing area, in order to
  avoid recomputation.
\item Every new solution is combined with the previously available
  solutions. Some of these will be recovered from the memoization
  memory and others may simply be available if they are the last
  solution computed by some goal and thus the bindings are active.
\item If more solutions are needed, backward execution
  is performed in parallel again. Goals which were suspended resume
  where they suspended.
\end{itemize}

All this brings the necessity of saving and resuming execution states,
memoing and recovering answers quickly, combining previously existing
solutions with newly found solutions, assigning agents to speculative
computations only if there are no non-speculative computations
available, and managing computations which change from speculative to
non speculative.
%
%
%
Note that all parallel backtracking is speculative work, because we
might need just one more answer of the rightmost parallel goal, and
this is why backward execution is given less priority than forward
execution.
%
%
Note also that at any point in time we only have one active value for
each variable. While performing parallel backtracking we can change
the bindings which will be used in forward execution, but before
continuing with forward execution, all parallel goals have to suspend
to reinstall the bindings of the answer being combined.


\compressection
\section{An Execution Example}
\label{sec:execution-example}

We will illustrate our approach, and specially the interplay of
memoization and parallel backtracking in IAP execution with the
following program:


\begin{pcode}
      main(X, Y, Z, T) @\neck@ a(X, Y) & b(Z, T).
      a(X, Y) @\neck@ a1(X) & a2(Y).
      b(X, Y) @\neck@ b1(X) & b2(Y).@
\end{pcode}

\begin{figure}[tb]
  \subfigure[Time = 0.]{
    \label{fig:example1}
    \includegraphics[width=0.22\linewidth]{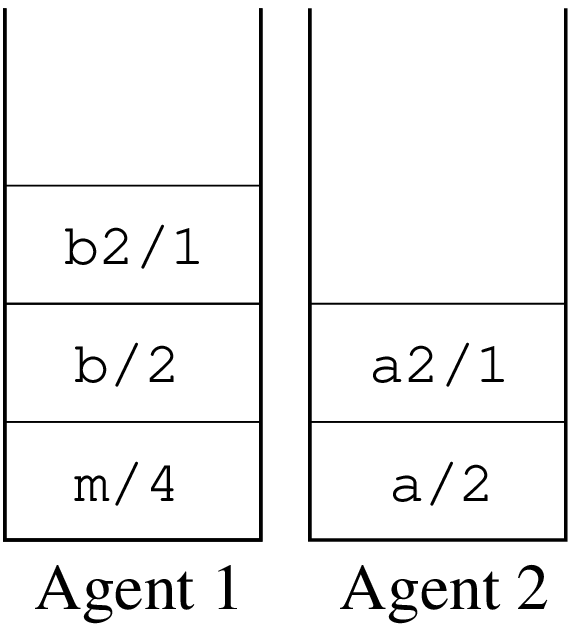}
  }
  \subfigure[Time = 3.]{
    \label{fig:example2}
    \includegraphics[width=0.22\linewidth]{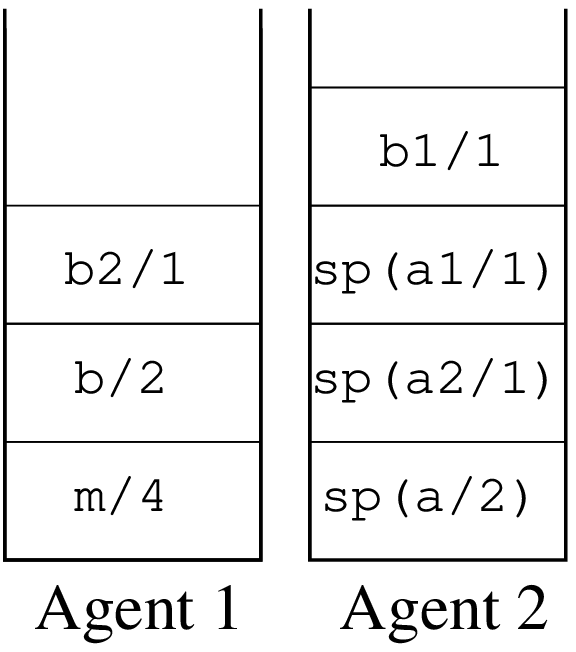}
  }
  \subfigure[Time = 4.]{
    \label{fig:example3}
    \includegraphics[width=0.22\linewidth]{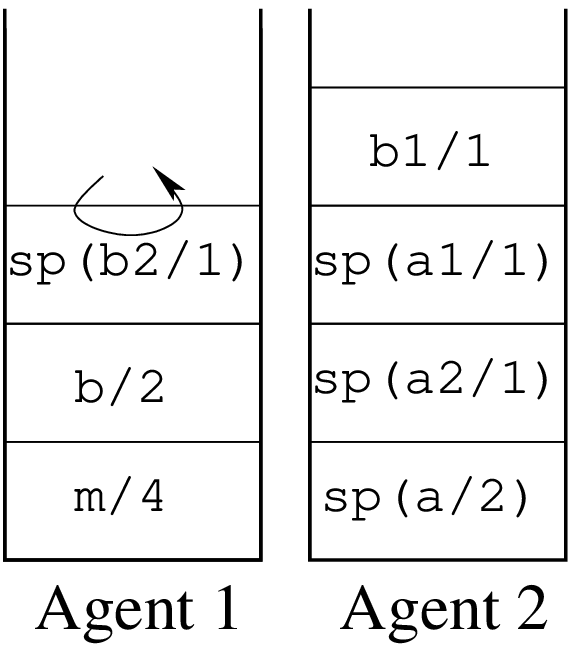}
  }
  \subfigure[Time = 6.]{
    \label{fig:example4}
    \includegraphics[width=0.22\linewidth]{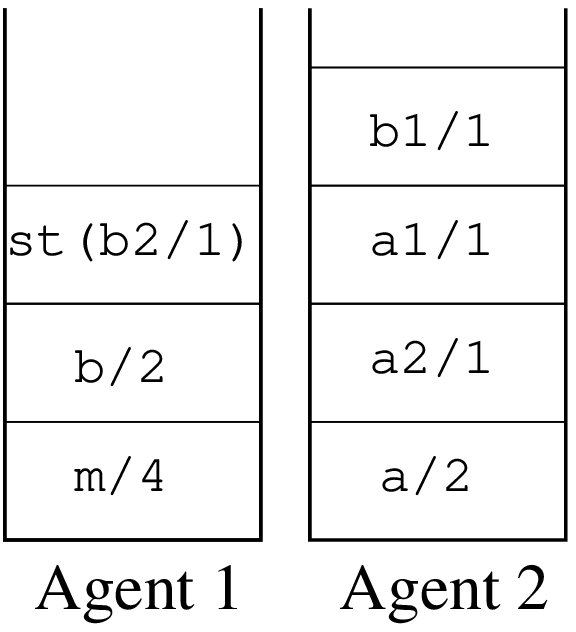}
  }
  \subfigure[Time = 16.]{
    \label{fig:example5}
    \includegraphics[width=0.22\linewidth]{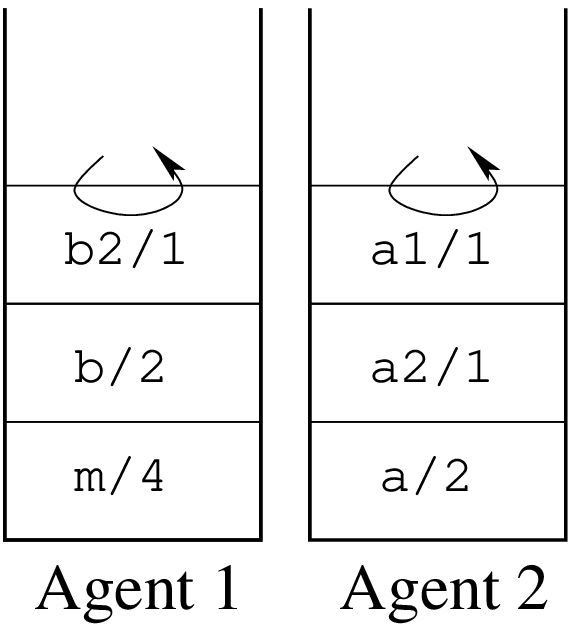}
  }
  \subfigure[Time = 23.]{
    \label{fig:example6}
    \includegraphics[width=0.22\linewidth]{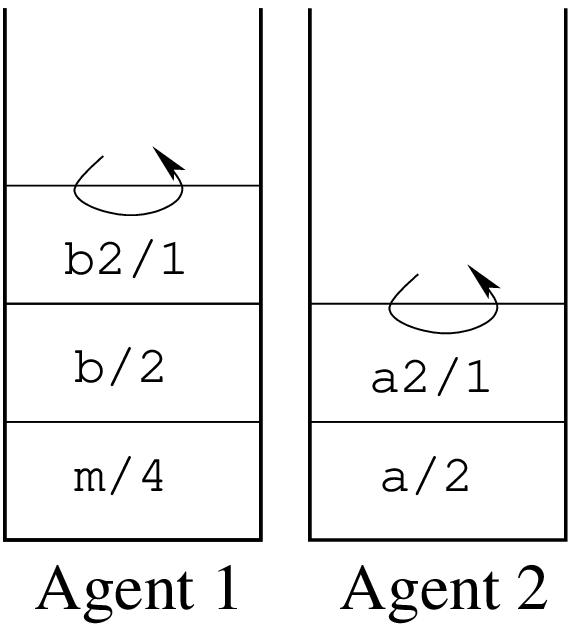}
  }
  \subfigure[Time = 29.]{
    \label{fig:example7}
    \includegraphics[width=0.22\linewidth]{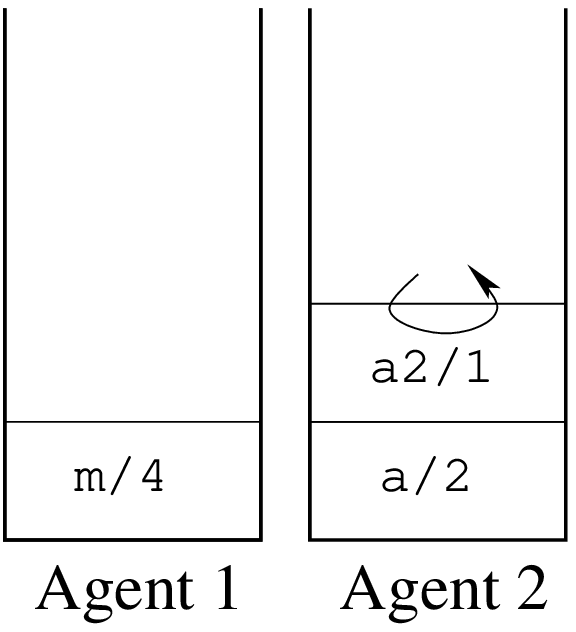}
  }
  \subfigure[Time = 36.]{
    \label{fig:example8}
    \includegraphics[width=0.22\linewidth]{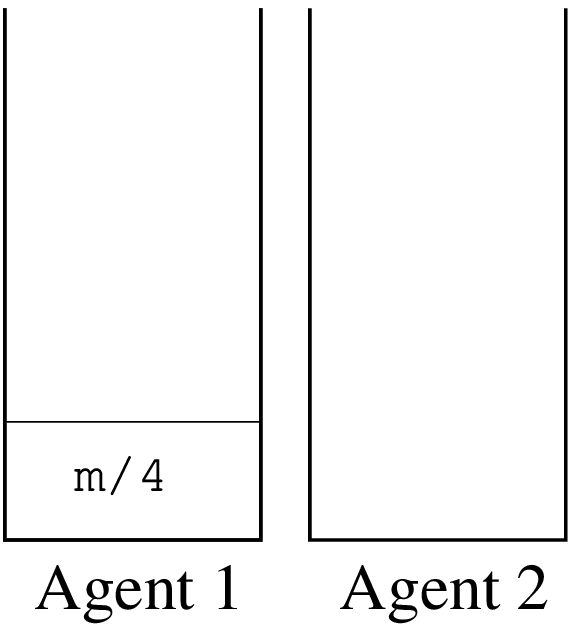}
  }
  \caption{Execution of \lstinline{main/4} with memoization of
    answers and parallel backtracking.}
  \label{fig:example}
\compressfigure
\end{figure}

We will assume that \lstinline{a1(X)}, \lstinline{a2(Y)},
\lstinline{b1(X)} and \lstinline{b2(Y)} have two answers each, which
take 1 and 7 seconds, 2 and 10 seconds, 3 and 13 seconds, and 4 and 25
seconds, respectively. We will also assume that there are no
dependencies 
among the variables in the literals of these clauses, and that the
cost of preparing and starting up parallel goals is
negligible. Finally, we will assume that there are two agents
available to execute these goals at the beginning of the execution of
the predicate \lstinline{main/4}. Figure~\ref{fig:example} summarizes
the evolution of the stack of each agent throughout the execution of
\lstinline{main/4} (abbreviated as \lstinline{m/4} in the figure).

Once the first agent starts the execution of \lstinline{main/4},
\lstinline{a/2} is published for parallel execution and
\lstinline{b/2} is executed locally. The second agent steals
\lstinline{a/2}, publishes \lstinline{a1/1} for parallel execution and
executes \lstinline{a2/1} locally, while the first agent marks
\lstinline{b1/1} as parallel and executes \lstinline{b2/1}. The
execution state can be seen in Figure~\ref{fig:example1}. When the
second agent finds the first answer for \lstinline{a2/1}, it marks
\lstinline{a2/1} to be executed in a speculative manner. However,
since \lstinline{a1/1} and \lstinline{b1/1} are still pending, the
second agent will start executing one of them instead. We will assume
it starts executing \lstinline{a1/1}.
Once it finds an answer, \lstinline{a1/1} is marked to be executed
speculatively. Since \lstinline{a2/1} is also marked as such, then the
entire predicate \lstinline{a/2} can be configured to be executed
speculatively. However, the second agent will now execute
\lstinline{b1/1} since it is pending and has higher priority than
speculative execution (Figure~\ref{fig:example2}).

Figure~\ref{fig:example3} shows the execution state when the first
agent finds an answer for \lstinline{b2/1}. In this case, since there is
no other parallel goal to execute, the first agent starts the
execution of \lstinline{b2/1} speculatively, until the second agent
finishes the execution of \lstinline{b1/1}. When that happens, the first
agent suspends the execution of \lstinline{b2/1} and the first answer of
\lstinline{main/4} is returned, as shown in Figure~\ref{fig:example4}.

In order to calculate the next answer of \lstinline{main/4}, both
agents will backtrack over \lstinline{b2/1} and \lstinline{b1/1},
respectively. Note that they would not be able to backtrack over other
subgoals because they are currently trapped.
Once the second agent
finds the second answer of \lstinline{b1/1}, the first agent suspends
the execution of \lstinline{b2/1} and returns the second answer of
\lstinline{main/4}, combining all the existing answers of its
literals.

In order to obtain the next answer of \lstinline{main/4}, the first
agent continues with the execution of \lstinline{b2/1}, and the second
agent fails the execution of \lstinline{b1/1} and starts computing the
next answer of \lstinline{a1/1}, since that goal has now been freed,
as shown in Figure~\ref{fig:example5}. Whenever the answer of
\lstinline{a1/1} is completed, shown in Figure~\ref{fig:example6}, the
execution of \lstinline{b2/1} is again suspended and a set of new
answers of \lstinline{main/4} involving the new answer for
\lstinline{a2/1} can be returned, again as a combination of the
already computed answers of its subgoals. To obtain the rest of the
answers of predicate \lstinline{main/4}, the first agent resumes the
execution of \lstinline{b2/1} and the second agent starts calculating
a new answer of \lstinline{a2/1} (Figure~\ref{fig:example7}). The
first agent finds the answer of \lstinline{b2/1}, suspends the
execution of the second agent, and returns the new answers of
\lstinline{main/4}. Finally, Figure~\ref{fig:example8} shows how the
second agent continues with the execution of \lstinline{a2/1} in order
to obtain the rest of the answers of \lstinline{main/4}.

Note that in this example memoization of answers avoids having to
recompute expensive answers of parallel goals. Also note that all the
answers for each parallel literal could have been found separately
and then merged, producing a similar total execution time. However,
the computational time for the first answer would have been
drastically increased.

\compressection
\section{Memoization vs.\  Recomputation}
\label{sec:reusing-answers}

Classic IAP uses recomputation of answers: if we execute
\lstinline{a(X) & b(Y)}, the first answer of each goal is generated in
parallel.  On backtracking, \lstinline{b(Y)} generates additional
answers (one by one, sequentially) until it finitely fails.  Then, a
new answer for goal \lstinline{a(X)} is computed in parallel with the
recomputation of the first answer of \lstinline{b(Y)}.  Successive
answers are computed by backtracking again on \lstinline{b(Y)}, and
later on \lstinline{a(X)}.

However, since \lstinline{a(X)} and \lstinline{b(Y)} are independent,
the answers of goal \lstinline{b(Y)} will be the same in each
recomputation.  Consequently, it makes sense to store its bindings
after every answer is generated, and combine them with those from
\lstinline{a(X)} to avoid the recomputation of \lstinline{b(Y)}.
Memoing answers does not require having the bindings for these answers
on the stack; in fact they should be stashed away 
and reinstalled when necessary.  Therefore, when a new answer is
computed for \lstinline{a(X)}
the previously computed and memorized answers for
\lstinline{b(Y)} are restored and combined.


\compressection
\subsection{Answer Memoization}
\label{sec:-answers-memoization}

In comparison with
tabling~\cite{tamaki.iclp86-short,Warren92,chen96:tabled_evaluation},
which also saves goal answers, our scheme shows a number of
differences:
we assume that we start off with terminating programs (or that if the
original program is non-terminating in sequential Prolog, we do not
need to terminate), and therefore we do not need to take care of the
cases tabling has to: detecting repeated calls,\footnote{Detecting
  repeated calls requires traversing the arguments of a goal, which
  can be arbitrarily more costly than executing the goal itself: for
  example, consider taking a large list and returning just its first
  element, as in
  {\scriptsize\lstinline{first([X|_],X)}}.}  suspending / resuming consumers,
maintaining SCCs, etc.
We do not keep stored answers after a parallel call finitely fails:
answers for 
%
\lstinline{a(X) & b(Y)} are kept for only as long as the new bindings
for \lstinline{X} and \lstinline{Y} are reachable. In fact,
we can discard \emph{all} stored answers as soon as the parallel
conjunction continues after its last answer.  Additionally, we
restrict the visibility of the stored answers to the parallel
conjunction: if we have
 \lstinline{a(X) & b(Y), a(Z)},
 the calls to \lstinline{a(Z)} do not have access to the answers for
 \lstinline{a(X)}.  While this may lead to underusing the saved
 bindings, it greatly simplifies the implementation and reduces the
 associated overhead.
 Therefore we will not use the memoization machinery commonly found in
 tabling implementations~\cite{rama95:efficient_tabling}.

 Instead, we save a combination of trail and heap terms which capture
 all the bindings made by the execution of a goal, for which we need
 two slight changes: we push a choicepoint before the parallel goal
 execution, so that all bindings to variables which live before the
 parallel goal execution will be recorded, and we modify the trail
 code to always trail variables which are not in the agent's
 WAM.\footnote{This introduces a slight overhead which we have
   measured at around 1\%.}  This ensures that all variable bindings
 we need to save are recorded on the trail.

 Therefore what we need to save are the variables pointed from the
 trail segment corresponding to the execution of the parallel goal
 (where the bindings to its free variables are recorded) and the terms
 pointed to by these variables. These terms are only saved if they
 live in the heap segment which starts after the execution of the
 parallel goal, since if they live below that point they existed
 before the parallel goal was executed and they are unaffected by
 backtracking.
 Note that bindings to variables which were created \emph{within} the
 execution of the parallel goal and which are not reachable from the
 argument variables do not have to be recorded, as they are not
 visible outside the scope of the parallel goal
 execution.\footnote{Another possible optimization
  is to share bindings corresponding to common parts of the search
  tree of a parallel goal: if a new answer is generated by performing
  backtracking on, for example, the topmost choicepoint and the rest
  of the bindings generated by the goal are not changed, strictly
  speaking only these different bindings have to be saved to save the
  new answer, and not the whole section of trail and heap.}  

\begin{figure}[tb]
  \centering\includegraphics[width=0.60\linewidth]{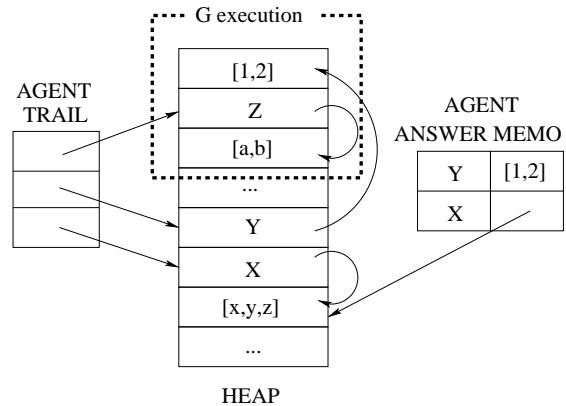}
  \caption{Snapshot of agent's stacks during answer memoization process.}
  \label{fig:answer_memo}
  \compressfigure
\end{figure}

Figure~\ref{fig:answer_memo} shows an example. \lstinline{G} is a
parallel goal whose execution unifies: \lstinline{X} with a list
existing before the execution of \lstinline{G}, \lstinline{Y} with a
list created by \lstinline{G}, and \lstinline{Z}, which was created by
\lstinline{G}, with a list also created by
\lstinline{G}. Consequently, we save those variables appearing in the
trail created by \lstinline{G} which are older than the execution of
\lstinline{G} (\lstinline{X} and \lstinline{Y}), and all the
structures hanging from them. \lstinline{[x,y,z]} is not copied
because is not affected by backtracking. The copy operation adjusts
pointers of variables in a way that is similar to what is done in tabling
implementations~\cite{rama95:efficient_tabling}. For example, if we
save a variable pointing to a subterm of \lstinline{[1,2]}, this
variable would now point to a subterm of the copy of
\lstinline{[1,2]}.

Note that this is at most the same amount of work as that of the
execution of the goal, because it consists of stashing away the
variables bound by the goal plus the structures \emph{created} by the
goal. The information related to the boundaries of the goal and
its answers is kept in a centralized per-conjunction data structure,
akin to a \emph{parcall frame}~\cite{ngc-and-prolog}. Similar
techniques are also used for the local stack.

Reinstalling an answer for a goal boils down to copying back to the
heap the terms that were previously saved and using the trail entries
to make the variables in the initial call point to the terms they were
bound to when the goal had finished.  Some of these variables point to
the terms just copied onto the heap and some will point to terms which
existed previously to the goal execution and which were therefore not
saved. In our example, \lstinline{[1,2]} is copied onto the heap and
unified with \lstinline{Y} and \lstinline{X} is unified with
\lstinline{[x,y,z]}, which was already living on the heap.
As mentioned before, while memoization certainly has a cost,
it can also provide by itself substantial speedups since it avoids
recomputations.  Since it is performed only on \emph{independent}
goals, the number of different solutions to keep does not grow
exponentially with the number of goals in a conjunction, but rather
only linearly.  This is an interesting case of synergy between two
different concepts (independence and memoization), which in principle
are orthogonal, but which happen to have a very positive mutual
interaction.

\compressection
\subsection{Combining Answers}
\label{sec:combining-answers}


When the last goal pending to generate an answer in a parallel
conjunction produces a solution, any sibling goals which were
speculatively working towards producing additional solutions 
have to suspend, reinstall the previously found answers, and combine
them to continue with forward execution.
A similar behavior is necessary when backtracking is performed over a parallel
conjunction and one of the goals which are being reexecuted in
parallel finds a new solution.
%
At this moment, the new answer is combined with all the previous
answers of the rest of the parallel goals.  For each parallel goal, if
it was not suspended when performing speculative backtracking, its
last answer is already on the execution environment ready to be
combined. Otherwise, its first answer is reinstalled on the heap
before continuing with forward execution.


When there is more than one possible answer combination (because some
parallel goals already found more than one answer), a \emph{ghost}
choice point is created.
%
This choicepoint has an ``artificial'' alternative which points to
code which takes care of retrieving saved answers and installing the
bindings. 
On backtracking, this code will produce the combinations of answers
triggered by the newly found answer (i.e., combinations already
produced are not repeated).  Note that this new answer may have been
produced by any goal in the conjunction, but we proceed by combining
from right to left.  The invariant here is that before producing a new
answer, all previous answer combinations have been produced, so we
only need to fix the bindings for the goal which produced the new
answer (say $g$) and successively installing the bindings for the
saved answers produced by the rest of the goals.

Therefore, we start by installing one by one the answers previously
produced by the rightmost goal.
When all solutions are exhausted, we move on to the next goal to the
left, install its next answer and then reinstall again one by one
the answers of the rightmost goal.  When all the combinations of
answers for these two goals are exhausted, we move on to the third
rightmost one, and so on ---but we skip goal $g$, because we only need
to combine its last answer since the previous ones were already
combined.


An additional optimization
is to update the heap top
pointer of the \emph{ghost} choice point to point to the current heap
top after copying terms from the memoization area to the heap, in
order to protect these terms from backtracking for a possible future
answer combination.  Consequently, when the second answer of the
second rightmost parallel goal is combined with all the answers of the
rightmost goal, the bindings of the answers of the rightmost goal do
not need to be copied on the heap again and then we only need to
untrail bindings from the last combined answer and redo bindings of
the answer being combined.  Finally, once the \emph{ghost} choice
point is eliminated, all these terms that were copied on the heap are
released.

One particular race situation needs to be considered. 
When a parallel goal generates a new
solution, 
other parallel goals may also find new answers before being suspended,
and thus some answers may be lost in the answer combination.  In order
to address this, 
our implementation maintains a pointer to
the last combined answer of each parallel goal in the parcall frame.
Therefore, if, e.g., two parallel goals, \lstinline{a/1} and
\lstinline{b/1}, 
have computed three answers each, but only two of them have been
combined, the third answer of \lstinline{a/1} would be combined with
the first two answers of \lstinline{b/1}, updating afterward its last
combined answer pointer to its third answer.  Once this is done, the
fact that \lstinline{b/1} has uncombined answers is detected before
performing 
backtracking, and the third answer of \lstinline{b/1} is combined with
all the computed answers of \lstinline{a/1} and, then, the last
combined answer of \lstinline{b(Y)} is updated to point to its last
answer.  Finally, when no goal is left with uncombined answers, the
answer combination operation fails.

\compressection
\section{Trapped Goals and Backtracking Order}
\label{sec:disordered-backtracking}

The classical, right-to-left backtracking order for IAP is known to
bring a number of challenges, among them the possibility of
\emph{trapped goals}: a goal on which backtracking has to be performed
becomes \emph{trapped} by another goal stacked on top of it.
Normal backtracking is therefore impossible.  Consider the following
example:

\begin{pcode}
  m(X,Y,Z) @\neck@ b(X,Y) & a(Z). 
  b(X,Y) @\neck@ a(X) &  a(Y).  
  a(1). a(2).@
\end{pcode} 

\begin{figure}[tb]
  \begin{minipage}[b]{0.67\linewidth}
    \includegraphics[width=1.00\linewidth]{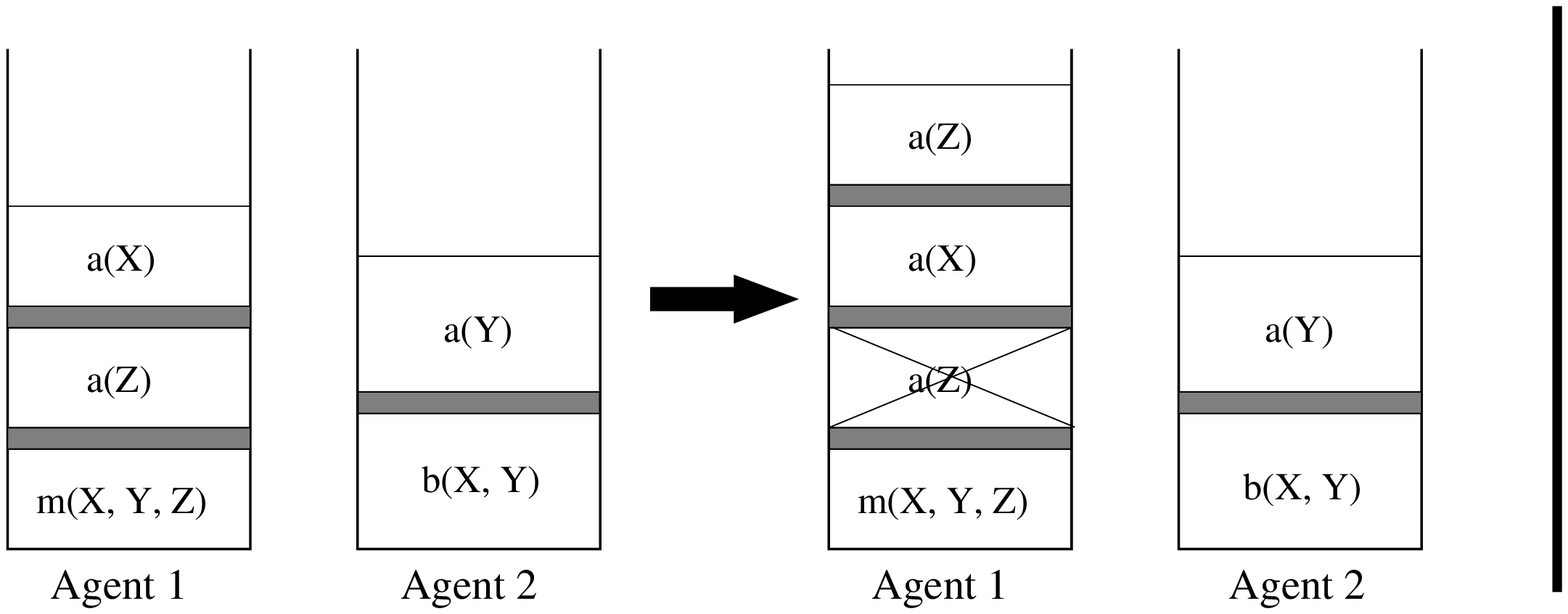}
    \caption{Execution of \lstinline{m/3}.}
  \label{fig:ordered_iap}
  \end{minipage}
  \hfill 
  \begin{minipage}[b]{0.28\linewidth}
    \includegraphics[width=1.0\linewidth]{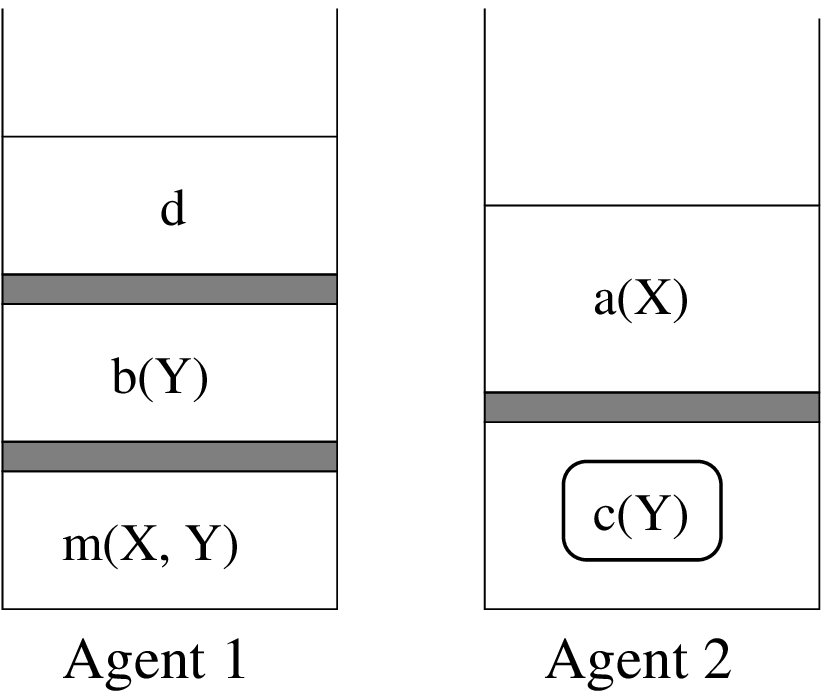}
    \caption{Execution of \lstinline{m/2}.}
  \label{fig:disordered_iap}
  \end{minipage}
\compressfigure
\end{figure}

Figure~\ref{fig:ordered_iap}\mnote{MCL: we need to mention what these sections mean and
  what the marker model is. DONE} shows a possible state of the
execution of predicate \lstinline{m/3} by two agents.  When the first
agent starts computing \lstinline{m/3}, \lstinline{b(X, Y)} and
\lstinline{a(Z)} are scheduled to be executed in parallel.  Assume
that \lstinline{a(Z)} is executed locally by the first agent and
\lstinline{b(X,Y)} is executed by the second agent. Then, the second
agent schedules \lstinline{a(X)} and \lstinline{a(Y)} to be executed
in parallel, which results in \lstinline{a(Y)} being locally executed by the
second agent and \lstinline{a(X)} executed by the first agent after
computing an answer for \lstinline{a(Z)}.
%
In order to obtain another answer for \lstinline{m/3}, right-to-left
backtracking requires computing additional answers for
\lstinline{a(Z)}, \lstinline{a(Y)}, and 
\lstinline{a(X)}, in that order.  However, \lstinline{a(Z)} cannot be directly
backtracked over since \lstinline{a(X)} is stacked on top of it: \lstinline{a(Z)} is a 
\emph{trapped goal}. 

Several solutions have been proposed for this problem.  One of
the original proposals uses \emph{continuation
  markers}~\cite{hermenegildo-phd-short,flexmem-europar96} to
\emph{skip} over stacked goals. 
This is, however, difficult to implement properly and needs to take
care of a large number of cases.  It can also leave unused
sections of memory (\emph{garbage slots}) which are either only
reclaimed when finally backtracking over the parallel goals, or
require quite delicate memory management.
A different solution~\cite{hlfullandpar-iclp2008} is to move the
execution 
of the trapped goal to the top of the stack.  This simplifies the
implementation somewhat, but it also leaves garbage slots in the
stacks.  

\compressection
\subsection{Out-of-Order Backtracking}
\label{sec:out-of-order-back}

Our approach does not follow the sequential backtracking order, to
reduce the likelihood of the appearance of trapped goals and garbage
slots.  The key idea is to allow backtracking (and therefore the
order of solutions) to dynamically adapt to the configuration of the
stacks.

As mentioned before, the obvious drawback of this approach is that it
may alter solution 
order with respect to sequential execution, and in an unpredictable
way.  However, we argue that in many cases this may  not be a high
price to pay, specially if the programmer is aware of it and can have
a choice. 
Programs where solution order matters, typically because of
efficiency, are likely to have dependencies between goals which would
anyway make them not amenable for IAP.
For independent goals we argue that allowing out-of-order backtracking
represents in some way a return to a simpler, more declarative
semantics that has the advantage of allowing higher efficiency in the
implementation of parallelism.

The alternative we propose herein consists of always backtracking over
the goal that is on top of the stack, without taking into account the
original goal execution order.\mnote{MH: But what if the goal on top
  is unrelated, i.e., from a different conjunction, than the one that
  we want to get another solution from?? Or it is related, or it is a
  soon of a related parallel goal which created a parcall frame. This
  parallel goal has to be performing backtracking too and it will
  finish at some point, although we can lose some parallelism here.}
For example, 
in the case of backward execution over predicate \lstinline{m/3} in
Figure~\ref{fig:ordered_iap}, both agents may be able to backtrack
over \lstinline{a(X)} and \lstinline{a(Y)}, without having to move the
execution of \lstinline{a(Z)}.  

\compressection
\subsection{First Answer Priority and Trapped goals}
\label{sec:memoing}

Out-of-order backtracking, combined with answer memoing not to lose
answer combinations, can avoid trapped goals if no priority is
given to any of the parallel goals, because there will always be a
backtrackable goal on the stack top to continue the execution of the
program. However, as mentioned before, we do impose a lightweight
notion of priority to first answers to preserve no-slowdown: 
backward execution 
of parallel goals that have not found any answer has more priority 
than backward execution of parallel goals which have already found an
answer.~\mnote{esta definicion no es exacta, pero no se como
  expresarla. No es que el objetivo paralelo haya o no haya encontrado
  una solucion, sino el objetivo que lo creo. En el ejemplo que sigue,
  c(Y) ya encontro una solucion, pero no b(Y), y por eso el
  backtracking sobre c(Y) es mas prioritario que sobre a(X), porque
  para encontrar la primera solucion de m(X,Y) necesitamos hacer
  backtracking sobre b(Y) y por tanto sobre c(Y). Si a(2) tarda mucho
  en computarse y no definimos esta prioridad, la primera solucion de
  m(X,Y) quedaria ineficientemente retardada sin necesidad.} 
Note that even using this very lax notion of priority, the possibility
of trapped goals returns, as illustrated in the following example:

\begin{pcode}
  m(X,Y) @\neck@ a(X) & b(Y). 
  b(Y) @\neck@ c(Y) &  d, e(Y).  
  a(1). a(2). c(1). c(2). d. e(2)@
\end{pcode} 

\noindent
Figure~\ref{fig:disordered_iap} shows a possible state of the
execution of predicate \lstinline{m/2} by two agents. The first agent
starts with the execution of predicate \lstinline{m/2} and publishes
\lstinline{a/1} and \lstinline{b/1} to be executed in parallel.  The
first agent starts with the execution of \lstinline{b/1} and marks
both \lstinline{c/1} and \lstinline{d/0} for parallel execution. The
second agent then executes \lstinline{c/1} while the first agent is
executing \lstinline{d/0}, and when the execution of \lstinline{c/1}
finishes then it computes an answer for \lstinline{a/1}. Once the
execution of goals \lstinline{c/1} and \lstinline{d/0} has finished,
\lstinline{e/1} is executed. However, this execution will fail because
\lstinline{c/1} already gave a different binding to variable
\lstinline{Y}. If the first answer is given priority, \lstinline{c/1}
should be backtracked before \lstinline{a/1}, but \lstinline{c/1} is
trapped by the execution of \lstinline{a/1}.  While 
this example shows that it is possible to have trapped goals with
out-of-order backtracking, we experimentally found that  the
  percentage of trapped goals  vs.\ remotely executed goals varies
  between 20\% and 60\% under right-to-left backtracking and it is
  always 0\% under out-of-order backtracking,
thus allowing for a
simpler solution for the  
problem without degrading the performance of parallel execution.

Our approach is to perform stack reordering to create a new execution
state which is consistent, i.e., which could have been generated by a
sequential SLD execution. Consequently, the parallel scheduler is
greatly simplified since it does not have to manage trapped goals.  We
cannot present the algorithm due to space limitations, but a high-level
view follows:

\begin{enumerate}
\item Copy the choice point and trail section corresponding to the
  trapped goal to the top of the stacks (their original allocations
  become garbage).
\item Move down the choice point and trail section to remove the
  generated garbage slots.
\item Update the trail pointers of relocated choice points to the 
  reordered trail section.
\item Keep heap and local stack in the same location. Global and frame
  stack top pointers of the trapped goal choice points are updated to
  point to the actual top of global and frame stack. Consequently, the
  execution memory of the goals that were moved down the stack is
  protected from backtracking.
\end{enumerate}

\mnote{MCL: we show speedups, but not how often goals are trapped.
  Can we get numbers for that? There would be trapped goals only in
  qsort\_nd for seqback, not for parback. This benchmarks is not
  actually working under seqback.}

\compressection
\section{The Scheduler for the Parallel Backtracking IAP Engine}
\label{sec:prolog-scheduler}
 
Once we allow backward execution over any parallel goal on the top of
the stacks, we can perform backtracking over all of them in
parallel. Consequently, each time we perform backtracking over a
parallel conjunction, each of the parallel goals of the parallel
conjunction can start speculative backward execution.

As we mentioned earlier, the management of goals (when a goal is
available and can start, when it has to backtrack, when messages have
to be broadcast, etc.) is encoded in Prolog code which
interacts with the internals of the emulator.
%
Figure~\ref{fig:scheduler} shows a simplified version of such a
scheduler, which is executed when agents (a) look for new work to do
and (b) have to execute a parallel conjunction. Note that
locks are not shown in the algorithm.

\begin{figure}[t]
  \centering
  \begin{minipage}[b]{0.49\linewidth}
    \begin{lstlisting}
parcall_back(LGoals, NGoals) @\neck@
    fork(PF,NGoals,LGoals,[Handler|LHandler]),
    (
        goal_not_executed(Handler) ->
        call_local_goal(Handler,Goal)
    ;   
        true
    ),
    look_for_available_goal(LHandler),
    join(PF).

look_for_available_goal([])@ \neck@ !, true.
look_for_available_goal([Handler|LHandler]) @\neck@ 
    (
        goal_available(Handler) ->
        call_local_goal(Handler,Goal)
    ;   
        true
    ),
    look_for_available_goal(LHandler).
    \end{lstlisting}
  \end{minipage}
\hfill
  \begin{minipage}[b]{0.4\linewidth}
    \begin{lstlisting}
agent @\neck@ work, agent.
agent @\neck@ agent.

work @\neck@
    find_parallel_goal(Handler) ->
    ( 
        goal_not_executed(Handler) ->
            save_init_execution(Handler),
            call_parallel_goal(Handler)
	;
	    move_execution_top(Handler),
            fail
    )
  ;
    suspend, 
    work.
    \end{lstlisting}
  \end{minipage}
  \caption{Parallel backtracking Prolog code.} 
  \label{fig:scheduler}
\compressfigure
\end{figure}

\compressection
\subsection{Looking for Work}
\label{sec:looking-for}

Agents initially execute the \lstinline{agent/0} predicate, which
calls \lstinline{work/0} in an endless loop to search for a parallel
goal to execute, via the \lstinline{find_parallel_goal/1} primitive,
which defines the strategy of the
scheduler.  Available goals can be in four states: non-executed
parallel goals necessary for forward execution, backtrackable parallel
goals necessary for forward execution, non-executed parallel goals not
necessary for forward execution (because they were generated by goals
performing speculative work), and backtrackable parallel goals not
necessary for forward execution.  Different scheduling policies are
possible in order to impose preferences among these types of goals
(to, e.g., decide which non-necessary goal can be picked) but studying
them is outside the scope of this paper.


Once the agent finds a parallel goal to execute, it is prepared to
start execution in a clean environment.
For example, if the goal has to be backtracked over and it is
trapped,
  a primitive operation
\lstinline{move_execution_top/1} moves the execution segment of the
goal to the top of the stacks to ensure that the choice point to be
backtracked over is always on the top of the stack
(using the algorithm of Section~\ref{sec:disordered-backtracking}).
Also, the memoization of
the last answer found is performed at this time, if the execution of
the parallel goal was not suspended. 

If \lstinline{find_parallel_goal/1} fails (i.e., no handler is
returned), the agent suspends until some other agent publishes more
work.  \lstinline{call_parallel_goal/1} saves some registers
before starting the execution of the parallel goal, such as the
current trail and heap top, changes the state of the handler once
the execution has been completed, failed, or suspended, and saves some
registers after the execution of the parallel goal in order to manage
trapped goals and to release the execution of the publishing agent.

\compressection
\subsection{Executing Parallel Conjunctions}
\label{sec:exec-par-conj}

The parallel conjunction operator \lstinline{&/2} is
preprocessed and converted into \lstinline{parcall_back/2}, which is
the entry point of the scheduler, and which receives the list of goals
to execute in parallel (\lstinline{LGoals}) and the number of goals in
the list.
%
\lstinline{parcall_back/2} invokes first \lstinline{fork/4}, written
in C, which creates a \emph{handler} for each parallel goal in the
scope of the parcall frame containing information related to that
goal, makes goals available for other agents to pick up, resumes
suspended agents which can then steal some of the new 
available goals, and inserts a new choice point in order to release all
the data structures on failure.

If the first parallel goal has not been executed yet, it is scheduled
for local execution\mnote{MCL: in many cases (e.g., usual recursion)
  it is much better to execute the last (or, in general, the recursive
  one) in order to spawn parallel goals as fast as possible.} by
\lstinline{call_local_goal/2}, which performs housekeeping similar to
that of \lstinline{call_parallel_goal/1}.  It can be already executed
because\mnote{MCL: I don't understand this?} this parallel goal, which
is always executed locally, can fail on backtracking, but the rest of
the parallel goals could still be performing backtracking to compute
more answers.  In this case, the choice point of \lstinline{fork/4}
will succeed on backtracking to continue forward execution and to wait
for the completion of the remotely executed parallel goals to produce
more answer combinations.

Then, \lstinline{look_for_available_goal/1} executes locally parallel
goals which have not already been taken by another agent.  Finally,
\lstinline{join/1} waits for the completion of the execution of the
parallel goals, their failure, or their suspension before combining
all the answers. After all answers have been combined, the goals of the
parallel conjunction are activated to perform speculative backward
execution.

\compressection
\section{Suspension of Speculative Goals}
\label{sec:goal-suspension}



Stopping goals which are eagerly generating new solutions
may be necessary for both correctness and performance reasons.
The agent that determines that suspension is
necessary sends a suspension event to the rest of the agents that
stole any of the sibling parallel goals (accessible via the parcall
frame).  These events are checked in the WAM loop each time a new
predicate is called, using existing event-checking machinery shared
with attributed-variable handling (and therefore no additional
overhead is added).
When the execution has to suspend, the argument registers are saved on
the heap, and a new choice point is inserted onto the stack to protect
the current execution state.  This choice point contains only one
argument pointing to the saved registers in order to
reinstall them on resumption.  The alternative to be executed on
failure points to a special WAM instruction which reinstalls the
registers and jumps to the WAM code where the suspension was
performed, after releasing the heap section used to store the
argument registers.  Therefore, the result of failing over this choice
point is to resume the suspended execution at the point where it was
suspended.

After this choice point is inserted, goal execution needs to jump back
to the Prolog scheduler for parallel execution.
In order to jump to the appropriate point in the Prolog scheduler
(after \lstinline{call_parallel_goal/1} or
\lstinline{call_local_goal/2}), the WAM frame pointer is saved in the
handler of the parallel goal before calling
\lstinline{call_parallel_goal/1} or \lstinline{call_local_goal/2}.
After
suspension takes place, it is reinstalled as the current frame
pointer, the WAM's \emph{next instruction} pointer is updated to be
the one pointed to by this frame, and this WAM instruction is
dispatched.
The result is that the scheduler continues its execution as if the
parallel goal had succeeded.

Parallel goals to be suspended may in turn have other nested parallel
calls.  Suspension events are recursively sent by agents following the chain of
dependencies saved in the parcall frames, similarly to the \emph{fail}
messages in \&-Prolog~\cite{ngc-and-prolog}.



\compressection
\section{A Note on Deterministic Parallel Goals}
\label{sec:note-det-parallel-goals}

The machinery we have presented can be greatly simplified when running
deterministic goals in parallel: answer memoization and answer
combination are not needed, and the scheduler
(Section~\ref{sec:prolog-scheduler}) can be simplified.
Knowing ahead of execution which goals are deterministic can be used
to statically select the best execution strategy.
%
%
However, some optimizations can be performed dynamically without
compiler support (e.g., if it is not available or imprecise).
For example, the \lstinline{move_execution_top/1} operation may decide
not to memoize the previous answer if there are no choice points
associated to the execution of the parallel goal, because that means
that at most one answer can be generated.\mnote{MCL: any other?}
By applying these dynamic optimizations, we have detected improvements
of up to a factor of two in the speedups of the execution of some
deterministic benchmarks.

\compressection
\section{Comparing Performance of IAP Models}
\label{sec:benchmarks}

We present here a comparison between a previous high-level
implementation of IAP~\cite{hlfullandpar-iclp2008} (which we
abbreviate as \textsf{seqback}) with our proposed implementation
(\textsf{parback}).  Both implementations are similar in nature and
have similar overheads (inherent to a high-level implementation), with
the obvious main difference being the support for parallel
backtracking and answer memoization in \textsf{parback}.
Both are implemented by modifying the standard
Ciao~\cite{ciao-reference-manual-1.13-short,hermenegildo11:ciao-design-tplp}
distribution. 
We will also comment on the relation with the very efficient IAP
implementation in~\cite{ngc-and-prolog} (abbreviated as
\textsf{\&-Prolog}) for deterministic benchmarks in order to evaluate
the overhead incurred by having part of the system expressed in
Prolog. 

We measured the performance results of both \textsf{parback} and
\textsf{seqback} on deterministic benchmarks, to determine the possible
overhead caused by adding the machinery to perform parallel
backtracking and answer memoization, and also of course on
non-deterministic benchmarks. 
The deterministic benchmarks used are the well-known Fibonacci series
(\emph{fibo}), matrix multiplication (\emph{mmat}) and QuickSort
(\emph{qsort}).
%
%
\emph{fibo} generates the 22$^{\mathrm{nd}}$ Fibonacci number
switching to a sequential implementation from the 12$^{\mathrm{th}}$
number downwards, \emph{mmat} uses 50x50 matrices and \emph{qsort} is
the version which uses \textsf{append/3} sorting a list of 10000
numbers.  The GC suffix means task granularity
control~\cite{granularity-jsc} is used for lists of size 300 and
smaller.
 
The selected nondeterministic benchmarks are \emph{checkfiles},
\emph{illumination}, and \emph{qsort\_nd}. \emph{checkfiles} receives a
list of files, each of which contains a list of file names which may
exist or not.
These lists are checked in parallel to find nonexistent files which
appear listed in all the initial files; these are enumerated on
backtracking.
\emph{illumination} receives an $N \times N$ board informing of
possible places for lights in a room.  It tries to place a light in
each of the columns, but lights in consecutive columns have to be
separated by a minimum distance.  The eligible positions in each
column are searched in parallel and position checking is implemented
with a pause of one second to represent task lengths.
\emph{qsort\_nd} is a QuickSort algorithm where list elements have
only a partial order.  \emph{checkfiles} and \emph{illumination} are
synthetic benchmarks which create 8 parallel goals and which exploit
memoization heavily.
\emph{qsort\_nd} is a more realistic benchmark which creates over one
thousand parallel goals.  All the benchmarks were 
parallelized using CiaoPP~\cite{ciaopp-sas03-journal-scp} and the
annotation algorithms described
in~\cite{annotators-jlp,daniel-phd,uudg-annotators-lopstr2007}.

Table~\ref{tab:speedup_results} shows the speedups obtained.
Performance results for \textsf{seqback} and \textsf{parback} were
obtained by averaging ten different runs for each of the benchmarks in
a Sun UltraSparc T2000 (a \emph{Niagara}) with 8 4-thread cores.  The
speedups shown in this table are calculated with respect to the
sequential execution of the original, unparallelized benchmark.
Therefore, the column tagged $1$ corresponds to the slowdown coming
from executing a parallel program on a single processor.  For
\textsf{\&-Prolog} we used the results in~\cite{ngc-and-prolog}. 
To complete the comparison, we note that one of the most efficient
Prolog systems, 
YAP Prolog~\cite{costa:yap-design-tplp}, 
very optimized for SPARC, is on these
benchmarks between 2.3 and 2.7 faster
than the execution of the
parallel versions of the programs on the parallel version of Ciao
using only one agent, but the parallel execution still outperforms
YAP.  Of course, YAP could in addition take advantage of parallel
execution.

For deterministic benchmarks, \textsf{parback$_{det}$} refers to the
implementation presented in this paper with improvements based on
determinacy information obtained from static
analysis~\cite{determ-lopstr04}.  For nondeterministic benchmarks we
show a comparison\mnote{Pablo, Amadeo: we do not actually say what we
  are comparing with.} of the performance results obtained both to generate
the first solution (\textsf{seqback$_{first}$} and
\textsf{parback$_{first}$}) and all the solutions
(\textsf{seqback$_{all}$} and \textsf{parback$_{all}$}).
Additionally, we also show speedups relative to the execution in
parallel with memoing in one agent (which should be similar to that
which could be obtained by executing sequentially with memoing) in
rows \textsf{pb\_rel$_{first}$} and \textsf{pb\_rel$_{all}$}.

 \begin{table}[t]
\resizebox{0.85\textwidth}{!}{%
  \begin{tabular}[b]{|c||l||c|c|c|c|c|c|c|c|}
    \cline{1-10}
    \raisebox{-2ex}[0em][-2ex]{\textbf{Benchmark}} &
    \raisebox{-2ex}[0em][-2ex]{\textbf{Approach}} &
    \multicolumn{8}{c|}{\textbf{Number of threads}} \\\cline{3-10}
    & & 1 & 2 & 3 & 4 & 5 & 6 & 7 & 8 \\\cline{1-10}

    \raisebox{-4ex}[0em][-4ex]{Fibo}
    & \textsf{\&-Prolog}           & 0.98 & 1.93 &  -   & 3.70 &  -   & 5.65 &  -   & 7.34 \\\cline{2-10}
    & \textsf{seqback}           & 0.95 & 1.89 & 2.80 & 3.70 & 4.61 & 5.36 & 6.23 & 6.96 \\\cline{2-10}
    & \textsf{parback}             & 0.95 & 1.88 & 2.78 & 3.69 & 4.60 & 5.33 & 6.21 & 6.94 \\\cline{2-10}
    & \textsf{parback$_{det}$}      & 0.96 & 1.91 & 2.83 & 3.74 & 4.65 & 5.41 & 6.28 & 7.04 \\\cline{1-10} 

    \raisebox{-4ex}[0em][-4ex]{QSort}
    & \textsf{\&-Prolog}           & 1.00 & 1.92 &  -   & 3.03 &  -   & 3.89 &  -   & 4.65 \\\cline{2-10}
    & \textsf{seqback}           & 0.50 & 0.98 & 1.38 & 1.74 & 2.05 & 2.27 & 2.57 & 2.67 \\\cline{2-10}
    & \textsf{parback}             & 0.49 & 0.97 & 1.37 & 1.74 & 2.05 & 2.27 & 2.58 & 2.69 \\\cline{2-10}
    & \textsf{parback$_{det}$}      & 0.56 & 1.10 & 1.54 & 1.96 & 2.31 & 2.57 & 2.90 & 3.02 \\\cline{2-10}
    & \textsf{seqbackGC}         & 0.97 & 1.77 & 2.42 & 3.02 & 3.37 & 3.77 & 3.98 & 4.15 \\\cline{2-10}
    & \textsf{parbackGC}           & 0.97 & 1.76 & 2.41 & 3.00 & 3.34 & 3.74 & 3.94 & 4.12 \\\cline{2-10}
    & \textsf{parbackGC$_{det}$}    & 0.97 & 1.78 & 2.44 & 3.04 & 3.41 & 3.79 & 3.99 & 4.21 \\\cline{1-10}

    \raisebox{-4ex}[0em][-4ex]{MMat}
    & \textsf{\&-Prolog}           & 1.00 & 1.99 &  -   & 3.98 &  -   & 5.96 &  -   & 7.93 \\\cline{2-10}
    & \textsf{seqback}           & 0.78 & 1.55 & 2.28 & 2.99 & 3.67 & 4.29 & 4.91 & 5.55 \\\cline{2-10}
    & \textsf{parback}             & 0.76 & 1.52 & 2.25 & 2.95 & 3.60 & 4.22 & 4.83 & 5.45 \\\cline{2-10}
    & \textsf{parback$_{det}$}      & 0.80 & 1.60 & 2.38 & 3.01 & 3.79 & 4.55 & 5.19 & 5.87 \\\cline{1-10}

    \raisebox{-4ex}[0em][-4ex]{CheckFiles}
    & \textsf{seqback$_{first}$} & 0.99 & 1.09 & 1.11 & 1.12 & 1.12 & 1.12 & 1.13 & 1.13 
    \\\cline{2-10}
    & \textsf{seqback$_{all}$}   & 0.99 & 1.05 & 1.07 & 1.07 & 1.07 & 1.08 & 1.08 & 1.08 
    \\\cline{2-10}
    & \textsf{parback$_{first}$}   & 3917 & 8612 & 10604 & 17111 & 17101 & 17116 & 17134 & 44222 
    \\\cline{2-10}
    & \textsf{pb\_rel$_{first}$}   &  1.00 & 2.20 & 2.71 & 4.37 & 4.37 & 4.37 & 4.37 & 11.29    
    \\\cline{2-10}
    & \textsf{parback$_{all}$}     & 12915 & 23409 & 30545 & 45818 & 46912 & 46955 & 46932 & 89571 
    \\\cline{2-10}
    & \textsf{pb\_rel$_{all}$}     & 1.00 & 1.81 & 2.37 & 3.55 & 3.63 & 3.64 & 3.63 & 6.94    
    \\\cline{1-10}

    \raisebox{-4ex}[0em][-4ex]{Illumination}
    & \textsf{seqback$_{first}$}  & 1.00 & 1.37 & 1.55 & 1.56 & 1.56 & 1.61 & 1.67 & 1.67
    \\\cline{2-10} 
    & \textsf{seqback$_{all}$}    & 1.00 & 1.16 & 1.21 & 1.24 & 1.24 & 1.25 & 1.25 & 1.27 
    \\\cline{2-10} 
    & \textsf{parback$_{first}$}    & 1120 & 1725 & 2223 & 3380 & 3410 & 4028 & 4120 & 6910
    \\\cline{2-10}
    & \textsf{pb\_rel$_{first}$}    & 1.00 & 1.54 & 1.98 & 3.02 & 3.04 & 3.60 & 3.68 & 6.17 
    \\\cline{2-10} 
    & \textsf{parback$_{all}$}      & 8760 & 16420 & 20987 & 31818 & 31912 & 31888 & 31934 & 65314  
    \\\cline{2-10}
    & \textsf{pb\_rel$_{all}$}      & 1.00 & 1.87 & 2.40 & 3.63 & 3.64 & 3.64 & 3.65 & 7.46 
    \\\cline{1-10}

    \raisebox{-4ex}[0em][-4ex]{QSortND}
    & \textsf{seqback$_{first}$}  & 0.94 & 1.72 & 2.36 & 2.92 & 3.25 & 3.59 & 3.78 & 3.92
    \\\cline{2-10} 
    & \textsf{seqback$_{all}$}    & 0.91 & 0.96 & 0.98 & 0.99 & 0.99 & 1.00 & 1.00 & 1.00
    \\\cline{2-10} 
    & \textsf{parback$_{first}$}    & 0.94 & 1.72 & 2.35 & 2.91 & 3.24 & 3.57 & 3.76 & 3.91
    \\\cline{2-10} 
    & \textsf{parback$_{all}$}      & 4.29 & 6.27 & 8.30 & 9.90 & 10.5 & 10.9 & 11.1 & 11.3 
    \\\cline{2-10}
    & \textsf{pb\_rel$_{all}$}      & 1.00 & 1.46 & 1.93 & 2.31 & 2.45 & 2.54 & 2.59 & 2.64 
    \\\cline{1-10}
  \end{tabular}}\hspace*{8em}
  \caption{Comparison of speedups for several benchmarks and implementations.}
  \label{tab:speedup_results}
\compressfigure
\vspace{-1ex}
\end{table}

The speedups obtained in both high-level implementations are very
similar for the case of deterministic benchmarks.  Therefore, the
machinery necessary to perform parallel backtracking does not seem to
degrade the performance of deterministic programs.

Static optimizations bring improved performance, but in this case they
seem to be quite residual, partly thanks to the granularity control.
When comparing with \textsf{\&-Prolog} we of course suffer from the
overhead of executing partly at the Prolog level (especially in
\textsf{mmat} and \textsf{qsort} without granularity control), but
even in this case we think that our current implementation is
competitive enough.  It is important that to note that the
\textsf{\&-Prolog} speedups were measured in another architecture
(Sequent Symmetry), so the comparison can only be indicative.
However, the Sequents were very efficient and orthogonal
multiprocessors, probably better than the Niagara in terms of
obtaining speedups (even if obviously not in raw speed) since the bus
was comparatively faster in relation with processor speed.
%
This
can only make \textsf{\&-Prolog} (and similar systems) have smaller
speedups if run in parallel hardware.  Therefore, their speedup could
only get closer to ours in current architectures.

\textsf{parback} and \textsf{seqback} behavior is quite similar
in the case of \emph{qsort\_nd} when only the first answer is computed
because there is not backtracking here.


In the case of \emph{checkfiles} and \emph{illumination}, backtracking
is needed even to generate the first answer, and memoing plays a
more important role.
The implementation using parallel backtracking is therefore much
faster even in a single processor since recomputation is avoided.
If we compute the speedup relative to the parallel execution on one
processor (rows \textsf{pb\_rel$_{first}$} and \textsf{pb\_rel$_{all}$})
the speedups obtained by \textsf{parback}
follow the increment in the number of processors more closely ---with
some superlinear speedup which is normal when search does not follow,
as in our case, the same order as sequential execution--- which can be
traced to the increased amount of parallel backtracking.
In contrast, the speedups of \textsf{seqback} do not increase so much
since it performs essentially
sequential backtracking.

When all the answers are required, the differences are still clearer
because there is much backward execution.
This behavior also appears, to a lesser extent, in
\emph{qsort\_nd}. More in detail, the \textsf{parback} speedups are not
that good when looking for all the answers of \emph{qsort\_nd} because
the time for storing and combining answers is not negligible here.

Note that the \textsf{parback} speedups of \emph{checkfiles} and
\emph{illumination} stabilize between 4 and 7 processors.  This is so
because they generate exactly 8 parallel goals, and there is one
dangling goal to be finished.  In the case of \emph{checkfiles} we get
superlinear speedup because there are 8 lists of files to check.  With
8 processors the first answer can be obtained without traversing (on
backtracking) any of these lists.  This is not the case with 7
processors 
and so there is no superlinear behavior until we hit the
8 processor mark.  
Additionally, since backtracking is done in parallel, the way the
search tree is explored (and therefore how fast the first solution is
found) can change between executions.


\compressection
\section{Conclusions}
\label{sec:conclusions}

We have developed a parallel backtracking approach for independent
and-paral\-lelism which uses out-of-order backtracking and relies on
answer memoization to reuse and combine answers. We have shown that
the approach can bring interesting simplifications when compared to
previous approaches to the complex implementation of the backtracking
mechanism typical in these systems. 
We have also provided experimental results that show significant
improvements in the execution of non-deterministic parallel calls due
to the avoidance of having to recompute answers and due to the fact
that parallel goals can execute backward in parallel, which was a
limitation in previous similar implementations. This parallel system
may be used in
applications
with a
constraint-and-generate
structure in which checking the
restrictions after the search is finished does not add significant
computation, and a simple code transformation allows a sequential
program to be executed in parallel.\mnote{But in general moving the
  checking after the generation greatly decreases performance to start
  with.  The cases in which this is not so can be very scarce.}


\bibliographystyle{acmtrans}
\bibliography{../../../bibtex/clip/clip,../../../bibtex/clip/general}

\end{document}